\documentstyle[12pt, aaspp4]{article}

\def\leaderfill{\leaders\hbox to 1em{\hss . \hss}\hfill}
\def\eg           {{\it e.g.~}}
\def\etal         {{et al.~}}

\begin{document}
 
\def\today{\number\year\space \ifcase\month\or  January\or February\or
        March\or April\or May\or June\or July\or August\or
September\or
        October\or November\or December\fi\space \number\day}
\def\fraction#1/#2{\leavevmode\kern.1em
 \raise.5ex\hbox{\the\scriptfont0 #1}\kern-.1em
 /\kern-.15em\lower.25ex\hbox{\the\scriptfont0 #2}}
\def\spose#1{\hbox to 0pt{#1\hss}}
\def\simlt{\mathrel{\spose{\lower 3pt\hbox{$\mathchar''218$}}
     \raise 2.0pt\hbox{$\mathchar''13C$}}}
\def\simgt{\mathrel{\spose{\lower 3pt\hbox{$\mathchar''218$}}
     \raise 2.0pt\hbox{$\mathchar''13E$}}}

\title{The New Transiting Planet OGLE-TR-56b: Orbit and Atmosphere.}
\author{Dimitar D. Sasselov}
\affil{Harvard-Smithsonian Center for Astrophysics, 60 Garden St., Cambridge, 
MA 02138}

\begin{abstract} 
Motivated by the identification of the very close-in extrasolar giant planet
OGLE-TR-56b ($P=1.2^d$, $a=0.023$AU, $M_p=0.9~M_{\rm Jup}$), 
we explore implications of its
existence on problems of tidal dissipation, planet migration, and atmospheric
stability. 
The small orbit of OGLE-TR-56b makes the planet an interesting test
particle case for tidal dissipation in stellar convection zones.
We show that it favors prescriptions of suppressed convective eddy
viscosity. Precise timing of the transits of OGLE-TR-56b might place 
interesting constraints on stellar convection theory
if orbital period change is detected in the near future.
   Another interesting issue related to the new planet is the origin of its
close orbit. OGLE-TR-56b may be a survivor of Roche-lobe overflow
mass loss. Such a precarious mechanism to halt inward planet migration, as
first suggested by Trilling \etal (1998), requires a large initial mass
($M_i>3~M_{\rm Jup}$) and sufficiently prompt disappearance of the remaining
disk ($\leq 10^6$ yrs).
   Living in such close proximity to its parent star, OGLE-TR-56b should
have temperatures in 
the dayside hemisphere reach 1900~K and we find that only few condensates
could form clouds, if at all, under those conditions $-$ mostly Al$_2$O$_3$
and Fe($s$). We explore the issue of atmosphere evaporation and find that
the planet's atmosphere is stable over its lifetime. The large size of the 
planet ($R_p=1.3~\pm{0.15}~R_{\rm Jup}$) might also require additional energy,
similar to HD~209458b.
In all, OGLE-TR-56b is 
our best ``laboratory" case yet to study extremes of planet atmospheres and
orbit dynamics.

\end{abstract}

\section{Introduction} 
The recent identification of the transiting giant planet OGLE-TR-56b 
(Konacki \etal 2003), was made possible by the successful photometric
search for shallow transits by the OGLE-III team (Udalski \etal 2002a,b).
Its parent star appears normal, with solar-like characteristics 
(M=1.04~M$_{\odot}$, R=1.10~R$_{\odot}$); so does the
giant planet $-$ with $M_p=0.9~M_{\rm Jup}$ \& $R_p=1.3~R_{\rm Jup}$
it is remarkably similar to HD~209458b (the only other exoplanet with known
radius). What sets OGLE-TR-56b apart is its orbit $-$ only half as large
(at 0.023~AU) than the more typical orbit of HD~209458b (at 0.045~AU).

While there is nothing ``typical" about a gas giant planet with a 0.045~AU
orbit, the past 7 years have brought us a sizeable class of close-in giants
like HD~209458b. Starting with 51~Peg (Mayor \& Queloz 1995), these now
number 13 and show a curious ``pile-up" at $a\sim 0.04$AU (or $P=3-4^d$).
The significant numbers of close-in giants suggests that inward migration
in the protoplanetray disk is common, and
the ``pile-up" indicates that a ``parking" mechanism with a sharp and robust
cutoff must exist (Kuchner \& Lecar 2002). This is important to
planet formation in general, because the current incompatibility between
formation timescales and migration timescales is not well understood.

Along comes OGLE-TR-56b, well inside the ``pile-up" (with $P=1.2^d$),
at almost the closest possible stable orbit. This Jupiter-mass
planet is close enough to raise variable tides on its star. The shear
associated with them is presumed to be dissipated in the stellar convection
zone by turbulent (eddy) viscosity (Zahn 1977; Lecar, Wheeler, McKee 1976). 
A central unresolved issue has remained in this physical picture
(Goodman \& Oh 1997; Terquem \etal
1998): how does the tide couple to the convection turbulence, \eg 
whether the largest eddies contribute to the total viscosity when
the period of the tide is less than the eddy turnover time.
Choosing a side in this controversy can affect orbit decay estimates
for close-in giant planets, as was pointed out by Rasio \etal (1996),
and illustrated by the results of some recent work (Patzold \& Rauer 2002;
Jiang, Ip, Yeh 2002). More generally, other affected issues are tidal
circularization of binary stars, stellar pulsations and oscillations;
this might be yet another indication that the now dated picture of
stellar envelope convection needs changing. OGLE-TR-56b is massive
enough to raise tides on its star, yet not massive enough to synchronize
it $-$ hence it is an ideal test particle for studies of tidal dissipation in
the star's envelope. 

In this paper, we study tidal decay and orbital stability
for OGLE-TR-56b in \S 3 and \S 4, after describing in some detail the derived
parameters of the OGLE-TR-56 star-planet system from the observations in
Konacki \etal (2003). Then in \S 5 we discuss the
problem of halting inward migration; and finally, in \S 6
we turn to the state of the planet's atmosphere.

\section{The Star-Planet Parameters} 

The stellar parameters were derived by modelling the high-resolution
Keck-HIRES spectra with numerical model atmospheres, and the star's mass and
radius were computed from our stellar evolution
tracks (Cody \& Sasselov 2002). 
We use the OPAL equation of state (Rogers et al. 1996) and
the latest Livermore opacity tables (OPAL96).
The nuclear reaction rates are calculated according to Bahcall 
\& Pinsonneault (1995). The diffusion of hydrogen and oxygen is
treated after Thoul, Bahcall, \& Loeb (1994). The upper boundary is
an atmosphere with a temperature distribution from model atmosphere
integration and Kurucz (1992) opacities; for models very different from the
present Sun, the Eddington $T-\tau$ relation is used. Convection is
treated with the standard mixing-length prescription and the
Schwarzschild stability criterion.
We find that OGLE-TR-56 is very similar to
the Sun, with a temperature of $T_{\rm eff} \approx 5900~\pm 80$~K, and solar
abundance of metals. The star's mass (M=1.04$\pm$0.05~M$_{\odot}$) and
radius (R=1.10$\pm$0.10~R$_{\odot}$) make it a somewhat closer analog
to HD~209458 than to the Sun. Combining the stellar parameters
with the OGLE-III $I$-band photometry and the radial velocity
data, yields the planetary mass,
radius and mean density: 0.9$M_{\rm Jup}$, 1.3$R_{\rm Jup}$, and
0.5 g/cm$^3$ (Konacki \etal 2003). We estimate the age of the star to 
be $3~\pm 1$~Gyrs, which is also consistent with the somewhat faster than
Sun's rotation (v$_{\rm rot}\approx 3~km~s^{-1}$) derived from our spectra.

The basic parameters of the OGLE-TR-56 system are:
orbital $P_{\rm orb}=2\pi/\omega=1.21190(\pm 0.9 s)$d,
star spin $P_{\rm spin}=2\pi/\Omega=20(\pm 5)$d,
separation $a=0.0225$AU,
mass ratio $q=M_{\rm p}/M=8.3(\pm 3.1)\times 10^{-4}$, and
mass in the stellar convection zone $M_{\rm cz}=0.021(\pm 0.005)M_{\odot}$.

\section{Tidal Decay, Orbit Stability, \& the Nature of Stellar Convection}

We assume that $\omega$ and $\Omega$ are aligned; this is very likely
and is known to be true for HD~209458 (Queloz \etal 2000), as well as
is possible to measure by upcoming spectroscopic observations.
We also assume a zero orbital eccentricity, $e=0$, but will explore
the timescales for circularization.
Although the current observations do not constrain the eccentricity
of OGLE-TR-56b's orbit, the theoretical expectation (e.g.,
Goldreich \& Soter 1966; Rasio \etal 1996) is that the tides
{\em on the planet} would have circularized its orbit a long time
ago (at age $t<10^9$ yrs). This is despite the fact that synchronization of
the planet's spin occurs much faster ($10^6$ yrs), following which
the dissipation factor is much lower.

The OGLE-TR-56 system is then a case of a slowly rotating star experiencing
fast tides, $\omega>\Omega$. {\em In addition,} the quadrupolar tidal forcing
due to $\omega$ occurs with timescales much shorter than the correlation
time of the turbulence in the star's convection zone. Tidal dissipation is
then expected to be strongly suppressed by local theory. 
Just how much should the
turbulent viscosity be inhibited has been a topic of controversy (see
Goodman \& Oh 1997 and references therein); OGLE-TR-56 might shed some
light on it.

First we study the tide raised by the planet OGLE-TR-56b on its star.
It is a quadrupole tidal wave of amplitude/height $h=qR^4a^{-3}~ 
(\approx{7~km}$)
above the hydrostatic equipotential of the stellar surface of radius $R$.
Dissipative processes in the star, \eg turbulent viscosity in its
convection zone, cause this tidal bulge raised by the planet to lag behind
by an angle $\delta$ radians, because $\omega>\Omega$. The bulge, of
mass $\propto M(h/R)$, then
exerts a torque on the planet in a direction opposite to its orbital
motion and the planet's orbit decays.
The star, or at least its convection zone, is spun up.

The rate of energy dissipation determines the rate of orbital evolution
which we aim to derive. We follow the paradigm developed by Lecar \etal
(1976) and Zahn (1977) for stars with convection zones; for now we keep $e=0$. 
The planet induces tidal motions of order $(\omega-\Omega){h}$, a large-scale flow
which sets up shear inside the star. The mostly radial convective motions
inside the star tend to erase the shear and dissipate. This coupling between
the tide and the convection eddies determines the magnitude of the
lag angle $\delta$. The amplitude, $h$, and the lag angle, $\delta$, of
the tidal distortion determine the torque exerted on the planet,
$\dot{L}\propto GM_{\rm p}^2k{\rm sin}(2\delta){\frac{R^5}{a^6}}$, and
provide us with a general equation for the orbital decay:
$$
\dot{a}={\frac{3GM_{\rm p}}{\omega{R^2}}}(1+q)\biggl({\frac{R}{a}}\biggr)^7{\epsilon(k_i,\delta)},
$$
where the complex physics of the tide-convection coupling is swept inside
the $\epsilon(k,\delta)$, which is called tidal coefficient by Zahn (1977) $-$ a
complex function of apsidal motion constants, $k_i$ (a simple stellar
interior description), and the lag angle, $\delta$. Alternatively, in a
somewhat more simple and transparent derivation by Lecar \etal (1976),
the above equation remains the same, but with $k\delta$ instead of
$\epsilon(k,\delta)$. One could compare the response of the star to that
of a forced harmonic oscillator and relate the phase angle to a specific
dissipation function $Q=(\omega-\Omega)E_0/\dot{E}=1/2\delta$ (Murray \&
Dermott 1999). Here, $E_0$ is the energy stored in the tidal bulge, and
$\dot{E}$, the rate of viscous dissipation of energy. In essence, this is
another way to estimate $\delta$, which has been used with assumed values
for $Q$'s of solar-like stars (\eg Patzold \& Rauer 2002). Below we continue
with the physically transparent approach; basically 
$Q\propto GM/R{\omega}{\nu}_{\rm t}$ in the parameters we use here.
Either way, the orbital decay rate will be a sensitive function
of $a$ and $R$, while estimating $\epsilon(k,\delta)$ (or $\delta$
alone) will require a prescription for the tide-convection coupling.

The adopted physical picture is based on the mixing-length theory
of convection (MLT) and assumes a Kolmogorov cascade of turbulent eddies
(Goldreich \& Nicholson 1977; Zahn 1989). 
The oscillatory tidal large-scale shear is
dissipated by turbulence, where the correlation time of the energy-bearing
turbulent eddies is much longer than the shear period. The turbulent (eddy)
viscosity is introduced in the sense of Rayleigh-Benard incompressible
convection, ${\nu}_{\rm t}={\frac{1}{3}}vl\approx l^2/{{\tau}_c}$, for a 
low forcing frequency. Here, $l$ is the convective mixing length, $v$ is
the convective velocity, and ${\tau}_c$ is the convective timescale
(eddy turnover time), in terms of MLT. Assuming a Kolmogorov cascade
gives us the scalings for these quantities for eddies smaller than
the mixing length. For a high forcing fequency, as is the case of OGLE-TR-56,
momentum transport by the largest eddies, of scale $l$, is inhibited.
As already mentioned above, two competing prescriptions for the amount of
this inhibition have been proposed: a linear one (Zahn 1977, 1989):
$$
{\nu}_{\rm t}={\frac{1}{3}}vl~min\biggl[\biggl({\frac{P_{\rm orb}}{2{\tau}_c}}\biggr),1\biggr],
$$
and a quadratic one (Goldreich \& Keeley 1977; Goldreich \& Nicholson 1977):
$$
{\nu}_{\rm t}={\frac{1}{3}}vl~min\biggl[\biggl({\frac{P_{\rm orb}}{2{\pi}{\tau}_c}}\biggr)^2,1\biggr].
$$
The former is introduced in part because it appears to lessen the disagreement
with observed circularization periods of solar-type stellar binaries, and
based on the large difference between the spatial scales of the tidal and
covective velocity fields. The quadratic prescription is central to the
theory of excitation and damping of the solar p-modes by convection, and
is generally better justified theoretically (Goodman \& Oh 1997).

Now we are ready to write a more direct expression for the timescale
of orbital decay, ${\tau}_a$, following the above picture (\eg Phinney 1992;
Rasio \etal  1996):
$$
\frac{\dot{a}}{a}=\frac{1}{{\tau}_a}={\frac{f}{{\tau}_c}}{\frac{M_{\rm cz}}{M}}
q(1+q)\biggl({\frac{R}{a}}\biggr)^8,
$$
The numerical factor $f$, obtained by integrating the total turbulent
viscosity throughout the convection zone,
is the focus of the theoretical controversy on the
efficiency of eddy viscosity. While its value is of order unity 
for ${\tau}_c\ll P_{\rm orb}$ (Verbunt \& Phinney 1995), for the case relevant 
to all close-in extrasolar planets and especially OGLE-TR-56b,
${\tau}_c\gg P_{\rm orb}$, and with
quadratic suppression of ${\nu}_{\rm t}$ we have
$f=min[1,(P/2{\pi}{\tau}_c)^2]$.

\section{The Orbit of OGLE-TR-56b}

We shall make use of the expression for ${\tau}_a$ from the previous section
to evaluate the stability of the orbit of the new transiting planet
OGLE-TR-56b. First, we should note that both the linear and quadratic
theoretical prescriptions for eddy viscosity suppression are in conflict
with the observations of the circularization of main-sequence binary stars
(Latham \etal 1992; Goodman \& Oh 1997; Terquem \etal 1998). A mechanism
other than turbulent convection needs to be invoked, or the MLT description
which gives us ${\tau}_c$ could be to blame. Of course,
binary stars are not good test cases for the problem we investigate here,
because they synchronize each other very fast, while OGLE-TR-56b is not
capable to spin-up its star, as we shall see here. For binary stars,
both the circularization,
${\tau}_{\rm circ}$, and decay, ${\tau}_a$, timescales are much longer than the
synchronization timescales, and are comparable to each other
(Zahn 1977; Terquem \etal 1998): 
${\tau}_a=6(1+q){\tau}_{\rm circ}$.

In the absence of other mechanisms, the possibility should be explored that
the culprit might be the incorrect
topology of convection implied by MLT. For example,
numerical simulations of stellar envelope convection (\eg Stein \& Nordlund
1989; Ludwig, Frytag, Steffen 1999) indicate large convective velocities
in a generally weakened vertical heat transport $-$ maintaining similar
length scales. This corresponds to an increase in the Brunt-V\"as\"al\"a
frequency, $|N^2|$; and as a reminder: ${\tau}_c=1/|N^2|^{1/2}$. 
Terquem \etal (1998) have shown that increasing
$|N^2|$ can come close to resolving the stellar circularization problem,
without contradicting helioseismological data.
OGLE-TR-56b's orbit could be used to test this idea and physical picture
(Sasselov, in prep.).

One could then derive an expression for the orbital decay, ${\tau}_a$,
which is calibrated to fit the observations, assuming a mechanism like
the increased $|N^2|$. This can be done with the known values for the 
circularization of solar-type binaries (Terquem \etal 1998; Meibom \& Mathieu 2003): 
${\tau}_{\rm circ}=4$~Gyr for $P_{\rm orb}=12.4$~days. With the additional
approximation, that ${\tau}_a$ is accurately determined by the initial
position of the planet (because the torque increases as it spirals in),
we have:
$$
{\tau}_a= 2.763\times 10^{-4}\frac{(1+q)^{\frac{5}{3}}}{q}P_{\rm orb}^{\frac{13}{3}},
$$
where $P_{\rm orb}$ is in days.

Using our interior models for the star OGLE-TR-56, we derive the parameters
of its convection zone, including ${\tau}_c=18 (\pm 2)$ days. First, we
estimate the orbital decay timescale for OGLE-TR-56b with the observationally
calibrated formula above:
$$
{\tau}_a=0.77\times 10^{9} yrs,
$$
with an uncertainty of at least $0.5\times 10^{9} yrs$ (Figure 1). Thus such large
amounts of dissipation, as seemed to be requred by stellar binaries,
appear excluded by the planet's orbit. Approaches that do not allow
for any viscosity suppression, $f\approx 1$, (e.g., Patzold \&
Rauer 2002) are excluded by the existence of OGLE-TR-56b.
Tidal dissipation mechanisms with more effective source of viscosity ($f>1$),
such as caused by magnetic fields, are even more strongly excluded by
OGLE-TR-56b. However, already Rasio \etal (1996) had shown that these were
excluded also by 51~Peg~b.

On the other hand, the theoretical models of turbulent viscosity dissipation
give comfortable estimates for the orbital stability of OGLE-TR-56b; for the
linear prescription:
$$
{\tau}_a=14.3(\pm 2.1)\times 10^{9} yrs,
$$
and for the quadratic prescription, the orbital decay is insignificant,
with:
$$
{\tau}_a=4.2\times 10^{12} yrs.
$$
The results are summarized in Figure 1, and point to the real possibility
for detecting changes in the orbit of OGLE-TR-56b by precise timing of its
transits. The observationally calibrated orbital decay estimates would give
$\dot{P}_{\rm orb}\sim 2$~msec/yr, with a range between 0.1 and 5 msec/yr,
given the large uncertainties (the quadratic prescription is undetectable). 
This is precision which can be achieved with HST photometry of the transit 
light curve (for a high quality template and zero point) over a 2-year
interval and additional ground-based observations. The currently available
ground-based transits have provided only 0.9~sec accuracy of its orbital period,
but this can be improved significantly by further dedicated observations and
an HST template and zero point.

\section{Evidence for a Different Parking Mechanism}

As illustrated in Figure 1, OGLE-TR-56b stands in a class of its own,
well to the left of the currently observed "pile-up" of close-in extrasolar
giant planets. It is an obvious candidate for the class II planet model
suggested by Trilling \etal (1998) and elaborated further by Lin \etal
(2000) $-$ planets that migrate inwards in
a protoplanetary disk, lose some (but not all) of their mass via Roche-lobe
overflow, and survive in very small orbits.  While the planet is losing
mass the inward torque from the disk is balanced by the outward torque
from the fast spinning pre-main-sequence star and the conservation of angular 
momentum (apart
from reducing the planet's mass). Nevertheless, a fairly prompt disappearance
of the inward torquing disk ($t\leq 10^6$ yrs) is required, because planets
in the 3-5~M$_{\rm Jup}$ mass range will be consumed within that time.
This is due in part to the M-R relation characteristic for giant planets
and the resulting flat $\dot{M}(M_p)$ function. 
Notably, it is not necessary for the entire circumstellar disk to
disappear, but just the inner part of it, which contributes enough inward
torque density to the planet at its close orbit. In other words, an
enlargement of the central disk hole during those $\approx 10^6~yrs$, would
have the same effect. Inward migration is then halted, as discussed by
Kuchner \& Lecar (2002).

The "survival" orbits are bound
on the inside by the Roche-lobe overflow and orbital decay boundaries (Figure
1). On the outside, these orbits will be limited by the amount of outward
torque to the planet by the faster spinning pre-main-sequence star.
Typical stellar spin periods at that early age are often $\leq 1$~day,
i.e., a planet at the Roche-lobe overflow orbit ($P_{\rm orb}>1$~day will
experience outward torque (same expressions as in \S 3 \& 4 apply, but
with opposite sign). With the rates discussed above, the planet's inward
migration could be stalled, but even if the spin-orbit interaction
continues as an outward torque after the disappearance of the disk,
the planet will not be able to migrate significantly outwards.

Such a mechanism would then create a separate distribution of
close-in giant planets, peaked around $1-1.5$~day periods, with
clear gap separating them from the now well-established distribution
of close-in planets at $3-4$~days. The timing experiment suggested
above for OGLE-TR-56b could be capable to reveal the presence of
a comparable-mass planet in a further out orbit, which might have
created this unusually close orbit.

\section{Stellar Irradiation and Stability of the Atmosphere}

OGLE-TR-56b is so close to its star, that its atmospheric structure
is completely dominated by the large incident stellar irradiation.
Here we use the modeling techniques described by Seager \& Sasselov (2000)
to compute the temperature structure on the dayside hemisphere of
the planet. The incident flux is from the model atmosphere calculation
for OGLE-TR-56 described in \S 2. The result for the dayside hemisphere
resembles closely the T-P profiles for other close-in extrasolar planets
(Seager, Whitney \& Sasselov 2000), with a very shallow temperature gradient all
throughout, except that OGLE-TR-56b is very hot, at $T_{\rm eff}$ = 1850 K,
even by ``hot Jupiter's" standards.

Given the very high temperature the issue of atmospheric stability to
evaporation is very relevant. The mean density of OGLE-TR-56b points to
hydrogen and helium as its main constituents. The loss of H and He from
a planet may occur by several means. To first order, Jeans "thermal
evaporation" gives a lower limit to the escape flux, even after corrections
for radiation pressure on satellite particle orbits are added (Chamberlain
\& Hunten 1987):
$$
F_{\rm Jea}(r_c)=\frac{N(r_c)U}{2\sqrt{\pi}}e^{-(\frac{v_{esc}}{U})^2}
\biggl(\frac{v_{esc}^2}{U^2} +1\biggr) ~~~~~[atoms~cm^{-2} s^{-1}],
$$
where $U^2=2kT/m$, with $m$ being the particle mass. This evaporation occurs
from the exobase ($r_c$) of the planet's atmosphere, where the particle
density is $N(r_c)$. The conditions at the
exobase are determined by our model. Namely, with $v_{esc}=38(\pm 4)~km~s^{-1}$
at the exobase and thermal r.m.s. velocity for hydrogen of $\sim 7~km~s^{-1}$,
we do not find any evidence for fast
evaporation or the existence of a blow-off state. The hydrogen escape flux is of
order $\sim 10^3$~g.s$^{-1}$. The existence of a planetary  magnetic field
could lead to larger mass loss fluxes, but currently we have no such evidence
(with the possible exception of a recent H Lyman $\alpha$ detection of an
extended exosphere around HD~209458b by Vidal-Madjar \etal 2003).

As regards the deeper atmosphere,
the high temperatures and small temperature gradient make the
case for cloud formation (of Al$_2$O$_3$, Fe(s), and Ti$_3$O$_5$)
unclear.
The structure of OGLE-TR-56b's atmosphere appears especially conducive to
the formation of ``iron clouds", Fe(s) condensates, due to their stronger
pressure dependance, e.g., compared to the other high-temperature condensate
Al$_2$O$_3$. Such clouds are highly absorbing (Seager, Whitney, \& Sasselov
2000) and would reduce significantly the albedo of OGLE-TR-56b.
Atmospheric dynamics, e.g., as modeled by Cho \etal (2002) and Showman \&
Guillot (2002), may further lessen the chances of any cloud formation,
except in localized low-pressure systems.

\section{Conclusions} 

OGLE-TR-56b has an orbit which is close enough to its star to
allow a test of tidal dissipation of stellar convection zone. It
represents a test which is different from the currently available
stellar binary circularization constraint, because the planet is
too small to spin-up the star.
In analyzing the orbital and physical parameters of the new extremely
close-in and hot transiting giant planet OGLE-TR-56b, we find that:

(1) its orbit is stable to orbital decay, if quadratic or linear
suppression of turbulent eddy viscosity in the parent star is assumed;
the linear prescription gives orbital decay rates which might be
detectable by a precision timing experiments using the transits,
and if successful could settle a long-standing issue regarding dissipation
in stellar convection zones.

(2) its atmosphere is stable to evaporation, and may contain occassional
Fe(s) and Al$_2$O$_3$ clouds, but would generally have a low albedo.

OGLE-TR-56b appears to define a new class of giant extrasolar
planets, in terms of the mechanism by which it attained its current
small orbit. The extreme parameters of this planet allow for the first
time some interesting constraints.

\acknowledgements
We are grateful to R. Noyes, M. Lecar, S. Phinney, and P. Goldreich for reading the
manuscript and detailed comments, and thank M. Holman, Y. Wu,
and N. Murray for useful discussions.

\begin{figure}
\plotfiddle{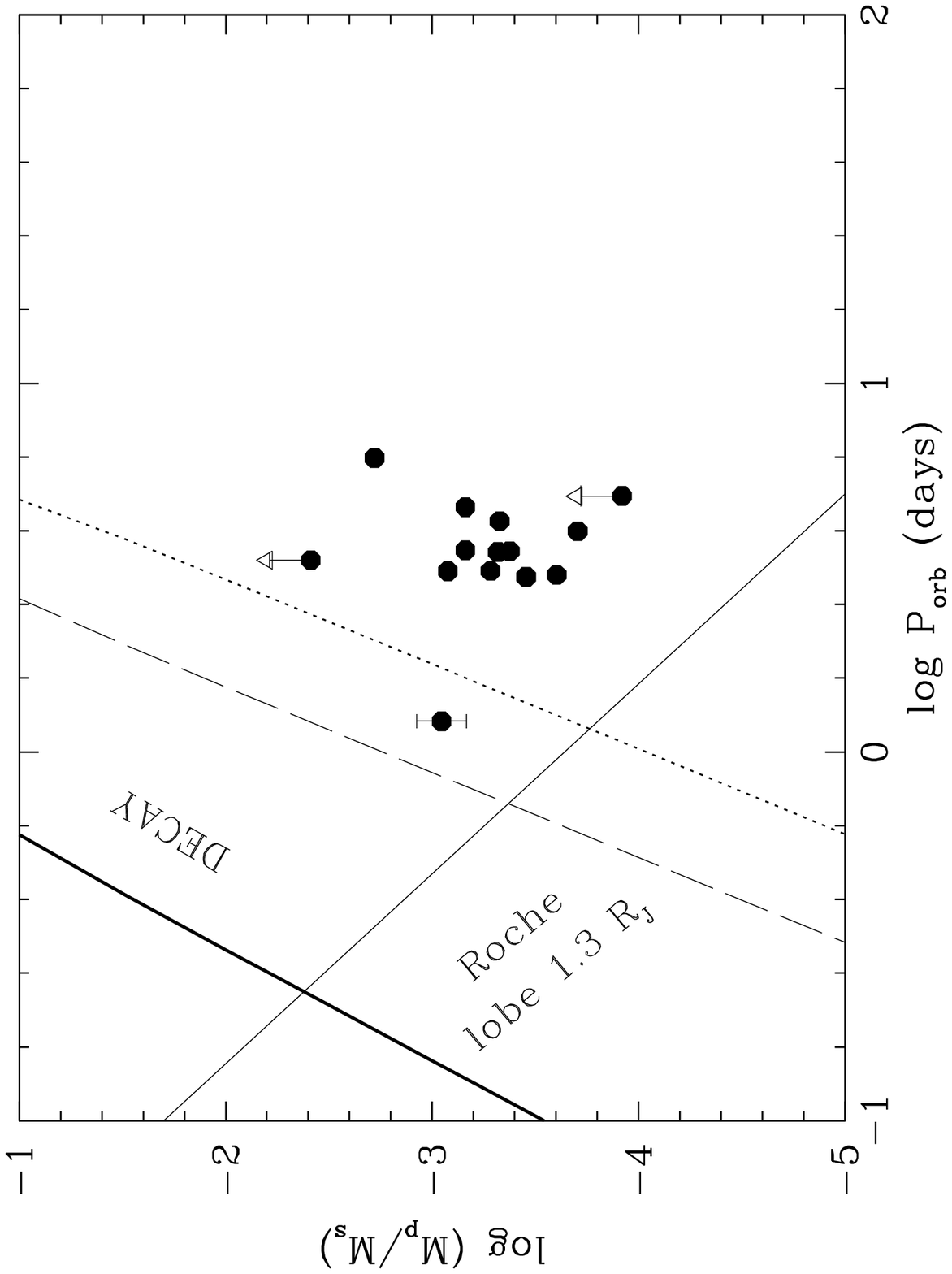}{15.0cm}{270}{90}{90}{-350}{500}
\caption{Orbital periods vs. masses of OGLE-TR-56b and all known close-in
giant planets. HD~209458b is the only other one with known mass; typical $sin~i$
uncertainties for the masses of the rest are shown by arrows for just two.
The areas to the left of the intersecting lines indicate
instabilities for a planet of mass $M_{\rm p}$ orbiting a solar-like
star like OGLE-TR-56 with $P_{\rm orb}$. The {\em dotted}, {\em dashed}, and
{\em heavy solid} lines mark parameter spaces beyond which the planet orbit
decays in 3~Gyrs (the age of OGLE-TR-56); they represent the observationally
calibrated (from stellar binaries), the theoretical linear suppression of eddy 
viscosity (${\nu}_{\rm t}$), and the theoretical quadratic suppression of
${\nu}_{\rm t}$ estimates, respectively.
The {\em thin solid} line marks Roche-lobe overflow for a planet radius
of 1.3$R_{\rm Jup}$ (OGLE-TR-56b's current size). The orbit of OGLE-TR-56b
is stable with the theoretical models for the stellar ${\nu}_{\rm t}$, but
would exhibit decay if larger dissipation is invoked, as required by the
circularization of solar-type binaries.}
\end{figure}

\end{document}